\documentclass[manuscript]{aastex62}
\usepackage{multirow}
\usepackage{amsmath}
\usepackage{longtable}
\usepackage{color}

\newcommand{\CCHb} {\mathrm{CCH}~(N=3-2,~J=5/2-3/2,~F=3-2)}
\newcommand{\CS} {\mathrm{CS}~(J=5-4)}
\newcommand{\SO} {\mathrm{SO}~(J_N=6_6-5_5)}
\newcommand{\vlsr}     {V_\mathrm{lsr}}

\newcommand{\pa} {\mathrm{P.A.}}

\newcommand{\h} {^{\mathrm{h}}}
\newcommand{\m} {^{\mathrm{m}}}
\newcommand{\s} {^{\mathrm{s}}}

\newcommand{\mJybeam}  {\mbox{mJy}~\mbox{beam}^{-1}}

\newcommand{\mJybeamkms}  {\mbox{mJy}~\mbox{beam}^{-1}~\mbox{km s}^{-1}}

\newcommand{\kms}	{\mbox{km s}^{-1}}

\newcommand{\au} {\mbox{au}}
\newcommand{\aukms} {\mbox{au km s}^{-1}}

\newcommand{\vlsrn}     {V_\mathrm{lsr,N}}
\newcommand{\vlsrs}     {V_\mathrm{lsr,S}}
\newcommand{\vrot}      {V_\mathrm{rot}}
\newcommand{\rf}      {R_\mathrm{footpoint}}

\shorttitle{Rotation in the IRAS 4C Outflow}
\shortauthors{Zhang et al.}

\begin{document}

\title{Rotation in the NGC 1333 IRAS 4C Outflow}

\author{Yichen Zhang}
\affiliation{Star and Planet Formation Laboratory, RIKEN Cluster for Pioneering Research, Wako, Saitama 351-0198, Japan}

\author{Aya E. Higuchi}
\affiliation{Star and Planet Formation Laboratory, RIKEN Cluster for Pioneering Research, Wako, Saitama 351-0198, Japan}

\author{Nami Sakai}
\affiliation{Star and Planet Formation Laboratory, RIKEN Cluster for Pioneering Research, Wako, Saitama 351-0198, Japan}

\author{Yoko Oya}
\affiliation{Department of Physics, The University of Tokyo, 7-3-1, Hongo, Bunkyo-ku, Tokyo 113- 0033, Japan}

\author{Ana L\'opez-Sepulcre}
\affiliation{Institut de Radioastronomie Millim\'etrique (IRAM), 300 rue de la Piscine, 38406 Saint- Martin-d'H\`eres, France}

\author{Muneaki Imai}
\affiliation{Department of Physics, The University of Tokyo, 7-3-1, Hongo, Bunkyo-ku, Tokyo 113- 0033, Japan}

\author{Takeshi Sakai}
\affiliation{Graduate School of Informatics and Engineering, The University of Electro-Communications, Chofu, Tokyo 182-8585, Japan}

\author{Yoshimasa Watanabe}
\affiliation{Division of Physics, Faculty of Pure and Applied Sciences, University of Tsukuba, Tsukuba, Ibaraki 305-8571, Japan}
\affiliation{Tomonaga Center for the History of the Universe, University of Tsukuba, Tsukuba, Ibaraki 305-8571, Japan}

\author{Cecilia Ceccarelli}
\affiliation{Universite de Grenoble Alpes, IPAG, F-38000 Grenoble, France}
\affiliation{CNRS, IPAG, F-38000 Grenoble, France}

\author{Bertrand Lefloch}
\affiliation{Universite de Grenoble Alpes, IPAG, F-38000 Grenoble, France}
\affiliation{CNRS, IPAG, F-38000 Grenoble, France}

\author{Satoshi Yamamoto}
\affiliation{Department of Physics, The University of Tokyo, 7-3-1, Hongo, Bunkyo-ku, Tokyo 113- 0033, Japan}

\begin{abstract}
We report molecular line observations of the NGC 1333 IRAS 4C outflow
in the Perseus Molecular Cloud with the Atacama Large
Millimeter/Submillimeter Array.
The CCH and CS emission reveal an outflow cavity structure with clear signatures of
rotation with respect to the outflow axis.
The rotation is detected from about 120 au up to about 1400 au above the envelope/disk mid-plane.
As the distance to the central source increases, 
the rotation velocity of the outflow decreases while the outflow radius increases,
which gives a flat specific angular momentum distribution along the outflow.
The mean specific angular momentum of the outflow is about 100 $\aukms$.
Based on reasonable assumptions on the outward velocity of the outflow and the protostar
mass, 
we estimate the range of outflow launching radii to be $5-15$ au.
Such a launching radius rules out that this outflow is launched as an X-wind,
but rather, it is more consistent to be a slow disk wind launched from relatively large radii on the disk.
The radius of the centrifugal barrier is roughly estimated, and 
the role of the centrifugal barrier in the outflow launching is 
discussed.
\end{abstract}


\section{Introduction}
\label{sec:intro}

Outflows play an important role in the star formation process, as they remove the angular momentum
from the accretion disks, allowing the material accreting to the protostars.
The outflows are believed to be driven magneto-centrifugally from the disk,
but the detailed mechanism of outflow launching is not completely understood.
Theoretical models differ in where the outflow is launched,
ranging from the innermost region of a disk close to the protostar (X-winds; e.g. \citealt[]{Shu00}),
to a wider range of radii on the disk surface (disk winds; e.g. \citealt[]{Konigl00}).
It is difficult to directly observe the launching area which is quite small
even in the case of disk winds.
However, the amount of angular momentum that an outflow transfers away from the disk
depends on the launching radius and disk size.
Therefore, by measuring the outflow rotation,
one can constrain the launching radii of the outflow.
Using such a method, recent observations have suggested that the low-velocity outflows with wide opening angles are launched from
about $2-25$ au on the disk, as the disk wind model predicts (e.g. \citealt[]{Bjerkeli16,Hirota17,Lee18}), 
and the high-velocity collimated jet may be launched from the innermost region of the disk, as an X-wind would suggest
(e.g. \citealt[]{Lee17}).
Some observation has also suggested that the axial jet and the wider slow outflow may be different parts of the same disk wind,
which is launched throughout the disk from the innermost region up to the outer disk (e.g. \citealt[]{Tabone17}).

In this paper, we report our recent observations
of the NGC 1333 IRAS 4C star-forming region with
the Atacama Large Millimeter/Submillimeter Array (ALMA),
which reveal a rotating molecular outflow.
Our target IRAS 4C is a low-mass protostar 
($L_\mathrm{bol}=0.7~L_\odot$; \citealt[]{Enoch09,Sadavoy14,Tobin16}) 
located in the NGC 1333 star-forming region in the
Perseus Molecular Cloud at a distance of 235 pc (\citealt[]{Hirota08}).
It is classified as a Class 0 source based on its bolometric temperature of $T_\mathrm{bol}=31~\mathrm{K}$ 
(\citealt[]{Enoch09,Sadavoy14,Young15,Tobin16}).
IRAS 4C lies 47$\arcsec$ northeast of the IRAS 4A star-forming region,
and 54$\arcsec$ north-northeast of the IRAS 4B star-forming region.
\citet[]{Koumpia16} pointed out that IRAS 4C may actually lie slightly in front of the IRAS 4A/B regions,
since absorption features seen in IRAS 4A/B coincide in velocity with the emission features of IRAS 4C.
This is also supported by the fact that IRAS 4A/B systems have a systemic velocity of about $+6.5~\kms$, while
IRAS 4C has a systemic velocity of about $+7.7~\kms$ (\citealt[]{Higuchi17}). 
Note that, IRAS 4B' (a.k.a IRAS 4BE or IRAS 4B II), 
the source 10$\arcsec$ east of IRAS 4B, has been sometimes called IRAS 4C as well. 
(e.g. \citealt[]{Looney00,Plunkett13}), but more often IRAS 4C is used for the source in this paper
(e.g. \citealt[]{Sandell01,Hatchell05,Tobin16}).

C$^{18}$O $(J=2-1)$ observations revealed a rotational structure in the north-south direction at
the center of IRAS 4C, but whether the rotation is Keplerian or not is not known due to the low
S/N  (\citealt[]{Tobin15}). {\it Spitzer} IRAC observations show an outflow cavity structure highlighted
by scattered light and shocked emission to the east side of the central source,
with the west side being much fainter (Figure 19 of \citealt[]{Tobin15}).
The east side is therefore inferred to be the blue-shifted side,
because the blue-shifted outflow cavity tends to be brighter than the red-shifted cavity in NIR and MIR
since it is less extincted. 
Despite the outflow cavity structure seen in infrared, there was no clear evidence of
a molecular outflow from (sub-)mm observations.
Previous  $^{12}$CO observations either reported no detection of outflow emission
toward this source (\citealt[]{Plunkett13,Tobin15}) or 
only weak compact blue-shifted emission towards east of the continuum source (\citealt[]{Stephens18}).
One possible explanation for the weak $^{12}$CO outflow emission is that
the outflow lies close to the plane of sky, so that the low-velocity outflow emission 
is easily mixed with the emission of the ambient gas, especially for abundant species like $^{12}$CO. 
Indeed, the inclination of the source is estimated to be nearly edge-on (\citealt[]{Tobin15})
to about $25^\circ$ between the disk plane and the line of sight (\citealt[]{SeguraCox16}).
On the other hand, the $^{13}$CO $(J=2-1)$ emission reveal a compact ($\lesssim2\arcsec$) structure with 
the blue-shifted emission slightly offset to the east of the red-shifted emission, 
which was explained as a slow outflow (\citealt[]{Koumpia16}).
Here we report that the outflow cavity structure is clearly detected in the CCH and CS emissions,
with kinematics consistent with rotation with respect to the outflow axis.
This allows us to measure the angular momentum distribution in the outflow,
and further constrain its launching radii and launching mechanism.

\section{Observations}
\label{sec:obs}

This source was observed as a part of the Perseus ALMA Chemical Survey 
(PEACHES; ALMA project 2016.1.01501.S; PI: N. Sakai) for 37 protostellar sources in the Perseus Molecular Cloud.
The observations were carried out in ALMA band 6 for four times from Nov. 26 to Nov. 30, 2016.
40 to 42 antennas were used and the baselines range from 15 to 700 m. The total integration time per source is 10 min.
J0237+2848 was used as the bandpass calibrator, J0238+1636 was used as the flux calibrators,
and J0336+3218 was used as the phase calibrator.
The source was observed with single pointing, 
and the primary beam size (half power beam width) is 22.9$\arcsec$ at Band 6.
The molecular lines were observed with velocity resolutions of about 0.14 $\kms$.
In this paper, we focus on the CCH $(N=3-2,J=5/2-3/2)$ lines, 
as well as the $\CS$ and $\SO$ lines (see Table \ref{tab:lines}).
Two hyperfine lines ($F=3-2$ and $F=2-1$) of the CCH $(N=3-2,J=5/2-3/2)$ transition were detected.
The SiO $(J=6-5)$ line was also observed simultaneously, but not detected.
In addition, a spectral window with a bandwidth of 938 MHz was used to map the 1.3 mm continuum. 
The data were calibrated and imaged in CASA (\citealt[]{McMullin07}). 
Self-calibration to the phase and amplitude was performed using the continuum data 
after the normal calibration, and applied to the molecular line data.
The CLEAN algorithm was used to image the data. For the spectral data we defined
its own clean region for each channel, encircling the area with the brightest emission. 
Robust weighting with the robust parameter of 0.5 was used in the clean process.
The resulting synthesized beam of the continuum data is $0.65\arcsec \times 0.41\arcsec$  ($\pa=-6.8^\circ$),
corresponding to about $150~\au\times 100~\au$.
The synthesized beams of the molecular line data are summarized in Table \ref{tab:lines}.
The largest recoverable scale is about $4\arcsec$.
The continuum peak is derived to be at 
$(\alpha_{2000},\delta_{2000})=(3\h29\m13\s.55$, $+31^\circ13\arcmin58\arcsec.11)$.

\section{Results}
\label{sec:result}

\subsection{Distribution}
\label{sec:distribution}

\begin{figure}
\begin{center}
\includegraphics[width=0.7\textwidth]{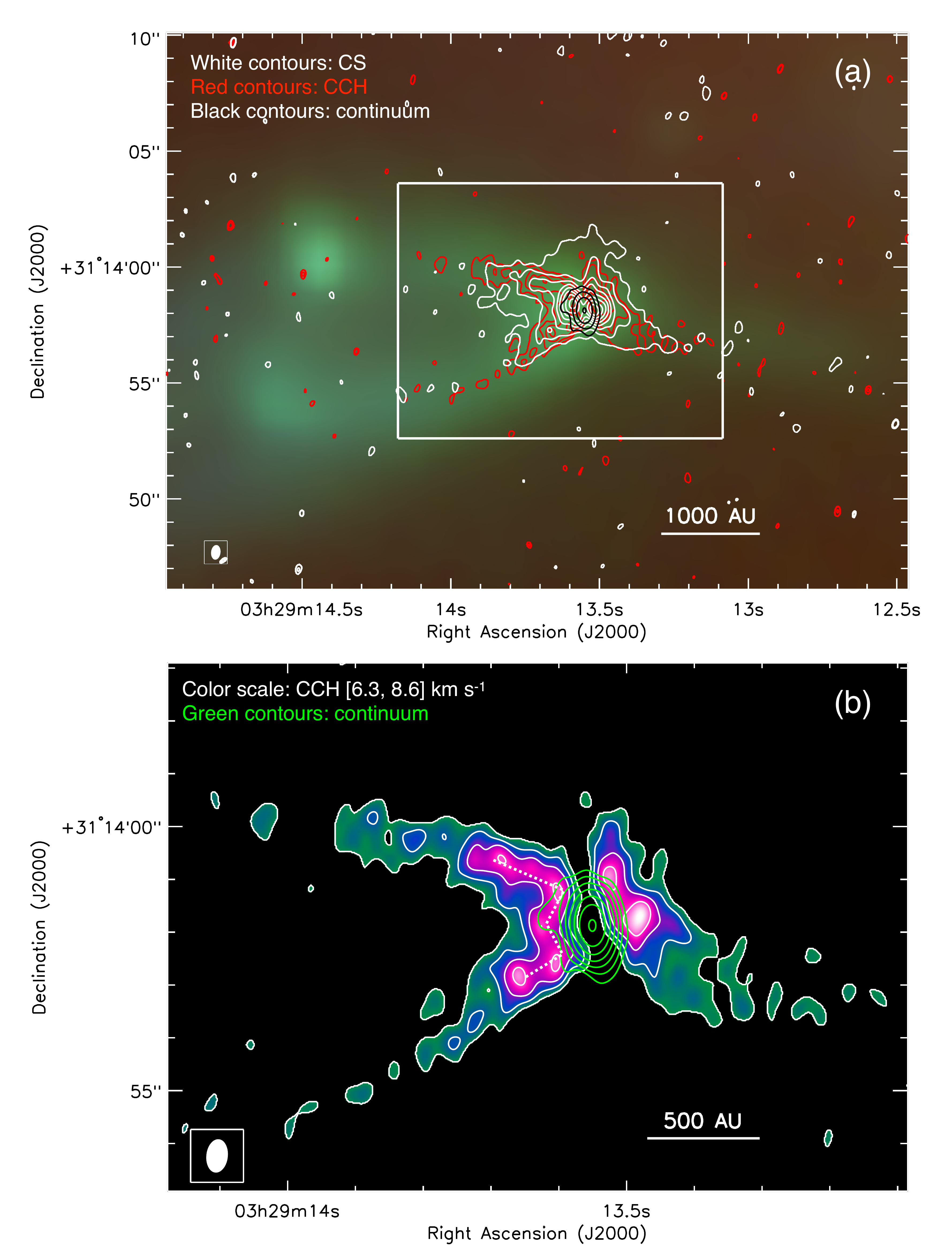}\\
\caption{{\bf (a):} The integrated emission maps of $\CCHb$ (red contours) and $\CS$ (white contours),
overlaid with the 1.3 mm continuum map in black contours.
The CCH emission is integrated from $\vlsr=6.3$ to 8.6 $\kms$,
and the $\CS$ emission is integrated from $\vlsr=6$ to 9.5 $\kms$.
The CCH contours start from $3\sigma$ and have intervals of $3\sigma$ with $1\sigma=5.2~\mJybeamkms$.
The CS contours start from $3\sigma$ and have intervals of $6\sigma$ with $1\sigma=7.4~\mJybeamkms$.
The continuum contours are at levels of $5\sigma$, $20\sigma$, $80\sigma$, 
and $320\sigma$, with $1\sigma=0.19~\mJybeam$.
The synthesized beam of the CCH data ($0.64\arcsec\times0.41\arcsec$) is shown in the bottom-left corner.
The back ground is the {\it Spitzer}-IRAC 3.6 $\mu$m, 4.5 $\mu$m, and 8.0 $\mu$m image.
{\bf (b)}: A zoom-in view of the CCH emission (color scale and white contours), overlaid with the continuum (green contours).
The continuum contours are at levels of $5\sigma$, $10\sigma$, $20\sigma$, $40\sigma$, $80\sigma$, $160\sigma$,
and $320\sigma$ ($1\sigma=0.19~\mJybeam$).}
\label{fig:intmap}
\end{center}
\end{figure}

Figure \ref{fig:intmap} shows the integrated maps of $\CCHb$ and $\CS$ emissions.
The CCH $(F=2-1)$ line has very similar distribution as the $F=3-2$ line shown in the figure,
but with slightly lower intensity.
The CCH emission shows a cavity structure with its base at the position
of the central source and extending toward the east. 
The morphology of this cavity structure is consistent with the outflow cavity seen in
mid-IR continuum.
The eastern lobe of the CCH outflow cavity appears to be highly axisymmetric.
The position angle of this outflow axis is 96$^\circ$, which 
is consistent with that estimated from the MIR images on a larger scale (\citealt[]{Tobin15}).
In contrast to the eastern side, the outflow cavity structure in the western side is incomplete.
The CCH emission is more concentrated toward the center and only the southern cavity is seen.
The asymmetry seen in the CCH line emission is very similar to that in MIR emissions,
suggesting that, in addition to the extinction, the material distribution may be indeed asymmetric
on the two sides of the outflows.
One possibility is that the western lobe
is affected by the red-shifted outflows from IRAS 4A.
However, it is hardly the case
if IRAS 4C is actually in front of IRAS 4A as suggested by \citet[]{Koumpia16}.
In any case, we focus our discussion on the eastern lobe of the outflow in this paper.

Close to the protostar, the continuum emission shows an elongated structure with its major axis perpendicular
to the outflow axis, which is consistent with the scenario that the continuum emission traces
an envelope/disk system around the protostar.
The size of the continuum source is determined to be $0.45\arcsec\times 0.21\arcsec$ ($106~\au\times 49~\au$)
with $\pa=18^\circ$.
The total continuum flux above $3\sigma$ within a radius of 2$\arcsec$ is 95 mJy, 
which corresponds to a mass of 0.061 $M_\odot$,
assuming a temperature of 30 K, dust opacity of \citealt[]{Ossenkopf94} 
($\kappa_\mathrm{1.3mm}=0.899~\mathrm{cm}^2~\mathrm{g}^{-1}$),
and the standard gas-to-dust mass ratio of 100.
Along the major axis of the continuum emission (i.e. the mid-plane of the envelope/disk system
assuming an edge-on inclination), there is a lack of CCH emission (see Figure \ref{fig:intmap}b). 
The CCH emission highlights only the surface of the disk-like structure.

Compared with the CCH emission, the CS emission is more concentrated toward the central source. 
Although it also traces the cavity structure of the outflow, a large amount of emission comes from inside 
the apparent cavity highlighted by the CCH emission. Close to the protostar, there are two emission peaks 
to the east and west of the continuum emission, i.e. above and below the mid-plane of the disk-like structure.
However, these emission peaks are closer to the mid-plane than the CCH emission.
Unlike the CCH emission, the CS emission also exists along the mid-plane and 
further extends to the south and north of the continuum emission.

\subsection{Outflow Cavities}
\label{sec:out}

\begin{figure}
\begin{center}
\includegraphics[width=\textwidth]{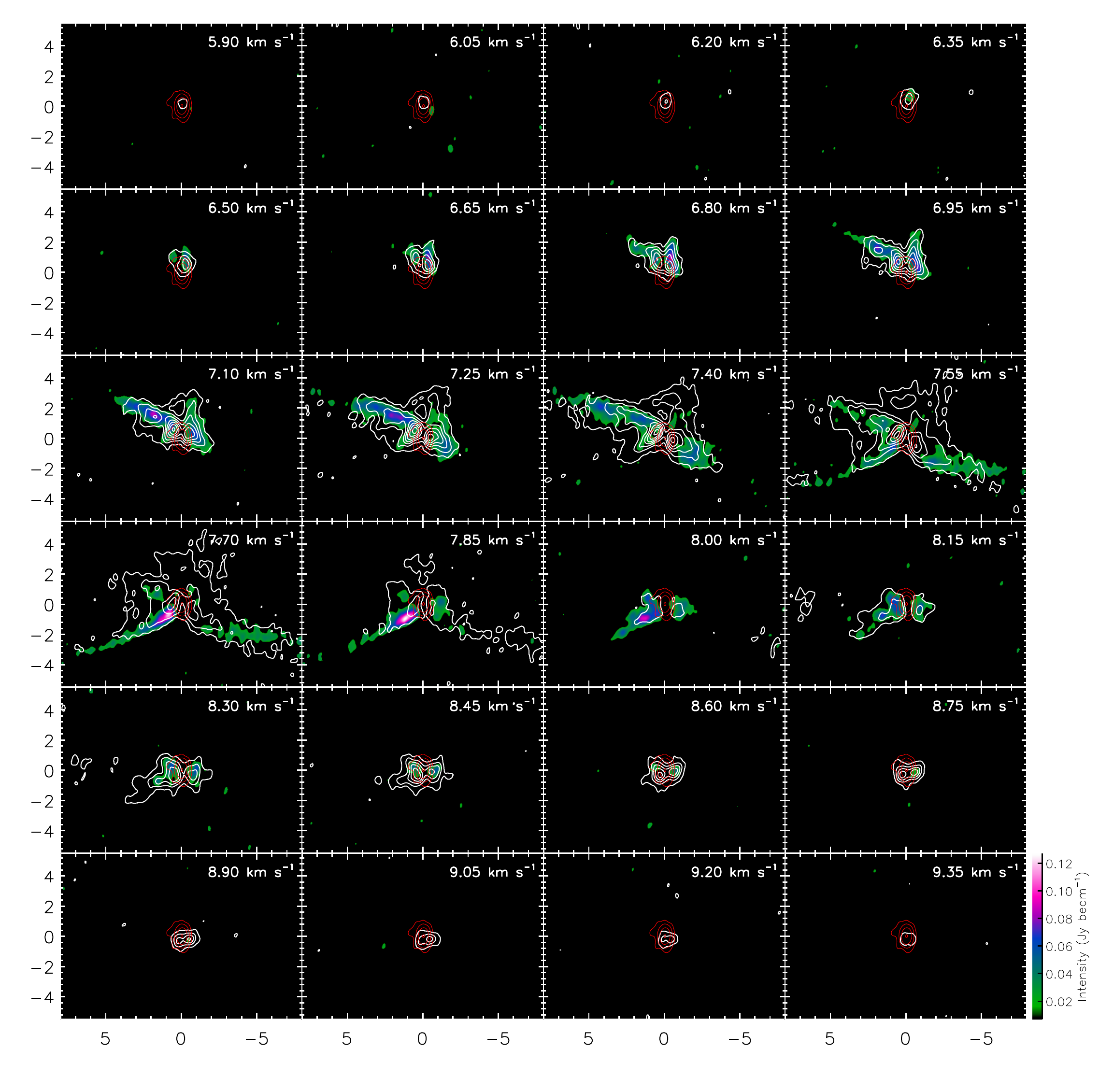}\\
\caption{The channel maps of the $\CCHb$ emission (color scale),
and the $\CS$ emission (white contours), overlaid with the 1.3 mm continuum emission in red contours.
The CS contours start from $3\sigma$ and have intervals of $6\sigma$ with $1\sigma=7.3~\mJybeam$.
The continuum contours are same as those in Figure \ref{fig:intmap}.
The channel width is 0.15 $\kms$, and 
the channel velocities are labeled at the upper-right corners of each panel.
The maps are rotated by 6$^\circ$ counterclockwise so that the outflow
axis is along the $x$ axis.
The origin of the position offsets is the continuum peak.}
\label{fig:channelmap}
\end{center}
\end{figure}

Figure \ref{fig:channelmap} shows the channel maps of the $\CCHb$ and $\CS$ emissions.
The outflow cavity structures are mostly detected within the velocity range of $6.8-8.3~\kms$.
There is little velocity difference between the eastern side and the western side of the outflow,
which, combined with the low velocity of the molecular outflow, is consistent with a nearly edge-on inclination.
However, there is clear velocity differences between the northern cavity wall of the outflow and the southern cavity wall,
which is best seen in the eastern lobe in the CCH emission.

In the CCH emission, at the blue-shifted velocity $\vlsr=6.8~\kms$, 
the northern cavity wall of the eastern lobe first appears
from the bottom of the cavity close to the central source. From $\vlsr=6.8~\kms$ to $7.4~\kms$,
this structure extends away from the center to the east, and reaches its maximum at $7.4~\kms$.
At $\vlsr=7.55~\kms$, both the northern and southern cavity walls are seen to have their maximum lengths.
We believe that this velocity corresponds to the central velocity of the eastern lobe.
At the red-shifted velocity $\vlsr=7.7~\kms$ to $8.15~\kms$, the northern cavity wall disappears, and the emission
in the southern cavity wall retreats from far away to the nearby region of the central source.
To sum up, in the eastern lobe, the low velocity CCH emission
traces full length of the outflow cavity while the high velocity emission only traces the base of the outflow cavity,
and the northern cavity wall is blue-shifted while the southern cavity wall is red-shifted.
Such a behavior can be easily explained by a rotating outflow, 
with lower rotation velocity far away from the central source at a larger radius from the outflow axis,
and higher rotation velocity closer to the central source at a smaller radius from the outflow axis.
Such distribution of rotation velocity is consistent with a constant specific angular momentum in the outflow.
The western outflow cavity shows similar behavior, but is not as clear as the eastern cavity due to its incomplete structure.
The CS emission also follows such a behavior. 
However, the rotation signature is not as prominent as the CCH emission,
because the CS emission is not only highlighting the projected boundary of the outflow cavity.
It is worth noting that the outflow cavity walls do not start from the position of the central source (i.e. the continuum peak), 
but instead from a position slightly offset from the central source.
The northern cavity wall seen in the CCH emission starts from about $1\arcsec$ to the north of the continuum peak
(best seen in the channel map of $\vlsr=7.1~\kms$), 
and the southern cavity wall starts from about $1\arcsec$ to the south of the continuum peak
(best seen in the CCH channel map of $\vlsr=8.0~\kms$).
Therefore the general morphology of the CCH emission is similar to a W-shape 
(see the white dotted line in Figure \ref{fig:intmap}b), 
instead of a V or U-shape usually seen in molecular outflows in larger scale observations.
This W-shape structure indicates that the outflow 
traced by CCH is not launched from the innermost region of the disk but a certain distance from the central star. 
However, the current resolution is not high enough for direct analysis of this launching region.

\begin{figure}
\begin{center}
\includegraphics[width=0.8\textwidth]{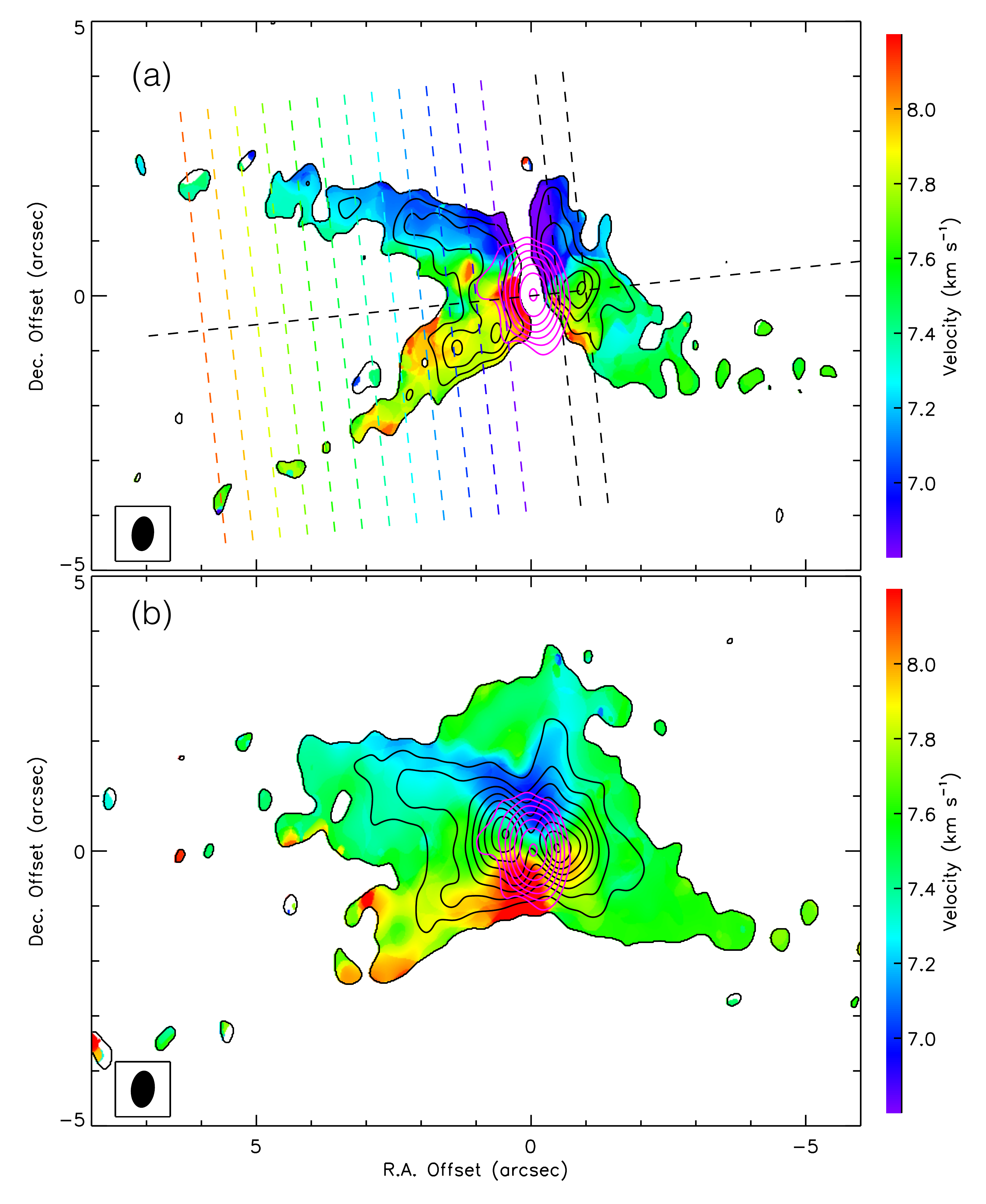}\\
\caption{{\bf (a):} The moment 0 map (black contours) and the moment 1 map (color scale)
of the $\CCHb$ emission, overlaid with the continuum emission in magenta contours. 
The east-west dashed line indicates the outflow axis.
The dashed lines perpendicular to the outflow axis indicate the cuts for
the position-velocity diagrams shown in Figures \ref{fig:pvdiagram} and \ref{fig:pvdiagram1}.
The colors of the cuts across the eastern lobe are used in Figure \ref{fig:pvdiagram1}.
{\bf (b):} Same as Panel (a), but for the $\CS$ emission.
In both panels, the moment 0 and continuum contours are the same as those in Figure \ref{fig:intmap}.}
\label{fig:momentmap}
\end{center}
\end{figure}

The velocity gradient is better seen in the moment 1 maps of CCH and CS emission in Figure \ref{fig:momentmap}.
It is clear that the outflow cavity structure seen in CCH, at least the eastern side,
has velocity structures highly symmetric to the outflow axis.
We will discuss the kinematic features in detail in \S\ref{sec:rotation}. 

\subsection{Envelope and Disk}
\label{sec:envelope}

As Figure \ref{fig:intmap} shows,
the CCH emission is missing on the mid-plane of the disk/envelope structure,
while the CS emission exists on the mid-plane and further extends to the south and north of the
continuum emission. Such differences can be seen in the high-velocity channels of Figure \ref{fig:channelmap}.
At high velocities of $\vlsr<6.8~\kms$ and $\vlsr>8.15~\kms$,
the CCH emission is concentrated in the vicinity of the central protostar, 
but always has some offsets with respect to the mid-plane
of the disk-like structure (major axis of the continuum).
The CS emission has similar behaviors, showing two peaks on the two sides
of the continuum structure, but at even higher velocities  ($\vlsr<6.2~\kms$ and $\vlsr>9.2~\kms$),
the CS emission is peaked on the mid-plane and closer to the central source.
These differences suggest that, although in this source the CS emission is mostly tracing the same material
as the CCH emission, it also traces the infalling material in the mid-plane which reaches closer
to the central source and has higher velocities from the rotation of the inner envelope or disk.
We will discuss such kinematic features of CS in more details in \S\ref{sec:disk}.
The different behaviors of the CCH and CS emissions indicate that 
there is a lack of CCH molecule on the envelope/disk mid-plane,
although the optical depth effect may also have contributed.
CCH is a reactive radical, and hence can be easily destroyed in a high density condition
of the mid-plane.
Meanwhile, the radiation from the protostar may further enhance the CCH formation
in the dense photodissociation region on the surface of the disk or envelope
(e.g. \citealt[]{Pety05,Oya17,Imai16}).
On the other hand, the CS molecule is relatively more stable
and may exist in the mid-plane toward the protostar.

\section{Discussions}
\label{sec:discussion}

\subsection{Outflow or Infall?}
\label{sec:outflow}

Before we go into the kinematics of the outflow,
we would like to argue that the CCH emission showing the outflow cavity structure actually
traces the outflowing gas along the outflow cavity, rather than the infalling gas surrounding the outflow cavity.
From the point of view of the CCH kinematics, the median velocity of the eastern CCH lobe is $7.5~\kms$ (see \S\ref{sec:angularmomentum}),
which is blue-shifted compared to the median velocities of the envelope traced by CS and SO emissions 
(see \S\ref{sec:disk}). This is consistent with the scenario that CCH traces outflowing gas,
since the blue-shifted $^{12}$CO outflow lobe is also seen towards the east of the central source (\citealt[]{Stephens18}).
In addition to this, we detect coherent rotation pattern from the inner region up to 
a large distance and high latitude (up to 1500 au at a polar angle of about $30^\circ$).
This is also consistent with an outflow, since the rotation pattern in the outflow can be well preserved 
due to the magnetic fields in the wind launching.
On the other hand, the rotation pattern of an infalling envelope gas with $\gtrsim 1000~\au$ may be affected by the turbulence,
the unsmooth and anisotropic distribution of the envelope material, or the change in direction of accretion.

From the point of view of chemical origin of CCH, there are mainly two mechanisms to form CCH in such protostellar sources.
First, CCH can be formed in photodissociation regions created by the UV radiation from the 
central protostar (e.g. \citealt[]{Pety05}), so it often traces the cavity walls of low velocity outflows 
(e.g. \citealt[]{Oya14,Oya18}). In those cases, the CCH emission shows kinematics consistent with outflow rather than infall.
Second, CCH is also formed in the envelope via warm carbon-chain chemistry (WCCC; \citealt[]{Sakai13}),
and in some cases it mainly traces the infalling envelope $\lesssim 500~\au$ (e.g. \citealt[]{Sakai14a}),
which also has rotation with conserved angular momentum.
But in such cases, the CCH emission has intensity peaks along the mid-plane of the inner envelope, 
although the excitation condition affects its intensity profile to some extent
(e.g. \citealt[]{Sakai14a,Sakai14,Oya15,Oya16,Oya17}).
Therefore, the CCH emission in this source, 
at least in the part distant from the disk mid-plane,
is most likely tracing the outflowing gas.
However, at the base of the outflow cavity, it is still possible that some fraction of the CCH emission
traces the gas on the surfaces of the disk and/or inner envelope. 

We note that such a scenario may be further complicated by the interferometric effect.
Compared with the observation of the same transitions of CCH by the IRAM 30 m telescope (\citealt[]{Higuchi17}),
our ALMA observation covers about 1/3 of the single dish fluxes.
The missing flux is most likely associated with the extended envelope which is resolved out in
the ALMA observation (the largest recoverable scale is about 1000 au in the observation).
However, the CCH emission in the extended envelope along the outflow cavity wall 
may not be resolved out but may be even emphasized
by the interferometric effect.
As mentioned above, the CCH gas motion in this source is outflowing,
and hence, even if the CCH molecules are formed in the extended envelope, they have already been entrained into the outflow.
Therefore we can still study outflow kinematics by the CCH emissions.

As partly discussed above, there are two possible phenomena, which the outflowing CCH gas traces.
The first is that the CCH traces the magneto-centrifugal wind launched from the disk.
The second is that the CCH traces an entrained outflow, composed of 
envelope material accelerated by the magneto-centrifugal wind.
For an entrained outflow, its momentum and angular momentum distributions
not only depend on the momentum and angular momentum distributions
in the magneto-centrifugal wind that accelerated it, but are also affected
by its original density distribution in the envelope,
especially the angular momentum distribution, since the rotation velocities of a 
magneto-centrifugal wind are usually much lower than its poloidal velocities.
Therefore, the fact that we detect clear and smooth distribution of rotation over a large distance
suggests that what we detected is unlikely to be an entrained outflow.
On the other hand, if the magneto-centrifugal wind is launched from
a relatively large radius (e.g. $\gtrsim 10~\au$) on the disk, its velocity can be at levels of a few tens of $\kms$ or lower.
Combined with the very low inclination, it is still consistent with the observed low velocities 
(see discussions in \S\ref{sec:out}).
For a distance of 200 au ($\sim1\arcsec$) above the disk and an outflow velocity of $10~\kms$, the 
dynamical timescale is about $10^2$ years, which is comparable to the timescale
to form molecules in the photodissociation regions (e.g. \citealt[]{Sakai13}). Therefore, the CCH molecules can form locally in the outflow.
However, close to the bottom of the outflow cavity, the CCH molecules may be formed on
the disk surface due to UV radiation and launched into the outflow.
Below, we will discuss the outflow rotation based on the assumption that
the CCH emission traces a magneto-centrifugal wind launched from the disk.

\subsection{Outflow Rotation}
\label{sec:rotation}

\subsubsection{Position-Velocity Diagrams of the Outflow}
\label{sec:pvdiagram}

\begin{figure}
\begin{center}
\includegraphics[width=0.85\textwidth]{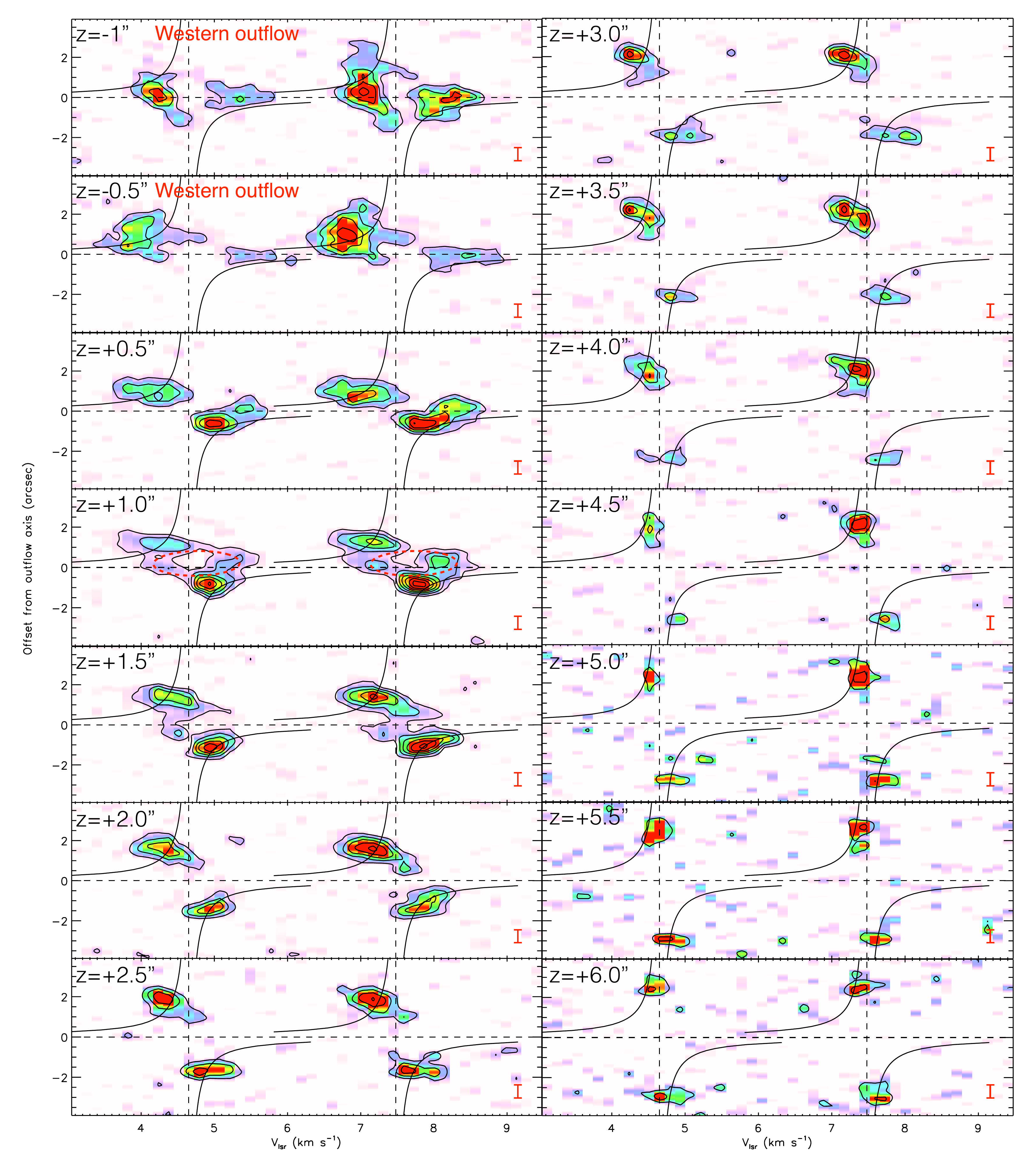}\\
\caption{The position-velocity diagrams of the CCH $(N=3-2,~J=5/2-3/2)$ emission along the cuts 
perpendicular to the outflow axis (shown in Figure \ref{fig:momentmap}).
The distances of these cuts to the central source are labeled at the
upper-left corners of each panel ($z>0$ for the eastern lobe and $z<0$ for the western lobe).
The contours start at $3\sigma$ and have intervals of $3\sigma$ with $1\sigma=5.2~\mJybeam$.
The color scale is relative to the maximum intensity in each panel.
The rest frequency of the $F=3-2$ hyperfine line is used,
so that the $F=2-1$ hyperfine line appears at a blue-shifted velocity. 
The black curves correspond to a constant angular momentum of 98 au $\kms$ (see \S\ref{sec:angularmomentum}).
The red ellipses in the panel of $z=1\arcsec$ are discussed in \S\ref{sec:pvdiagram}.
The red bar at the lower-right corner of each panel indicates the resolution beam size.}
\label{fig:pvdiagram}
\end{center}
\end{figure}

\begin{figure}
\begin{center}
\includegraphics[width=\textwidth]{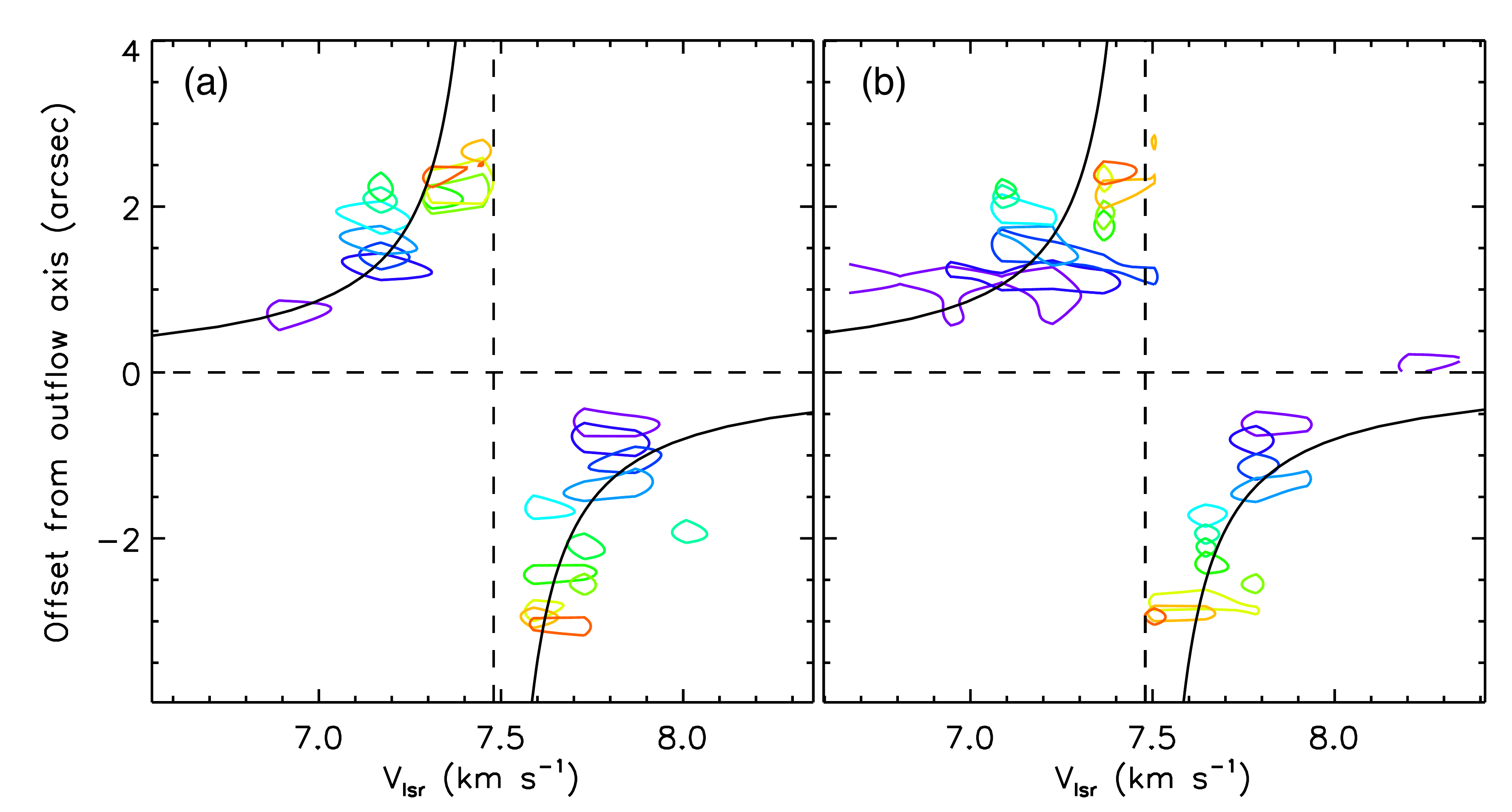}\\
\caption{The position-velocity diagram showing the emission peaks of
different panels of the eastern lobe in Figure \ref{fig:pvdiagram} in one panel.
Panel (a) is for the CCH $(F=3-2)$ emission, and Panel (b) is for the CCH $(F=2-1)$ emission.
The color corresponds to the different cuts shown in Figure \ref{fig:momentmap}.
The contours are at levels of 0.9 of the maximum intensity in each cut on each side.
The curves are the same as those in Figure \ref{fig:pvdiagram}.}
\label{fig:pvdiagram1}
\end{center}
\end{figure}

Figure \ref{fig:pvdiagram} shows the position-velocity (PV) diagrams of the CCH emission
in the eastern lobe and the base of the western lobe,
along cuts perpendicular to the outflow axis (the cuts are shown in Figure \ref{fig:momentmap}).
Here the outflow axis passes through the continuum peak
and has a position angle of 96$^\circ$ to the north.
Each cut has a width of 0.5$\arcsec$ and has a distance
to the central source ranging from $z=+0.5\arcsec$ to $z=+6\arcsec$ in the eastern lobe,
and $z=-0.5\arcsec$ and $-1\arcsec$ in the western lobe.
For the eastern lobe ($z>0$), in each panel, 
the velocities of the emission peaks associated with the northern wall (positive offsets) and
the southern wall (negative offsets) are symmetric with respect to a median velocity
which is about 7.5 $\kms$ (the vertical dashed lines).
This median velocity does not appear to change with $z$, indicating a nearly constant forwarding velocity
of the outflow (no acceleration or deceleration), or a nearly edge-on inclination.
As we discussed in \S\ref{sec:result}, 
the velocity difference between the northern and southern outflow cavity walls decreases with $z$, 
while the position offset between the northern and southern outflow cavity walls increases with $z$,
indicating rotation with nearly constant angular momentum.
A curve showing a constant angular momentum with respect to the outflow axis of 98 $\aukms$ (0.42 arcsec $\kms$)
is shown in each panel. Only the eastern lobe is used to determine such a value (see \S\ref{sec:angularmomentum}).
The emission peaks in each PV diagram in the eastern lobe ($z>0$) follow this curve quite well.
This is better illustrated by Figure \ref{fig:pvdiagram1}, 
where we plot the emission peaks (defined as $0.9~\times$ the peak intensities)
of all panels with $z>0$ of Figure \ref{fig:pvdiagram} in a single panel.
The emission in the western lobe ($z<0$) appears to be also consistent with it.

Besides the emission associated with the outflow cavity walls, 
at $0.5\arcsec\leq |z|\leq1.\arcsec$ (i.e. 120 au $\leq |z| \leq$ 240 au)
from the envelope/disk mid-plane on both sides, 
there are blue and red-shifted emission inside the apparent outflow cavity
around the position of axis (see Figure \ref{fig:pvdiagram}).
These emissions correspond to the center part of the W-shape of the integrated emission
seen in Figure \ref{fig:intmap}.
They can also be seen in the channel maps of $\vlsr=7.1$ and $8.0~\kms$ in Figure \ref{fig:channelmap}.
These emissions, together with the emission associated with the cavity walls, form
elliptical shapes in the PV diagrams (best seen in the panel of $z=+1\arcsec$, labelled by red dashed ellipses).
Such kinematic structure can be explained by both an expanding shell with rotation (e.g. \citealt[]{Hirota17}),
or infalling-rotating motion (e.g. \citealt[]{Sakai14}).
Since the PV slice is 1$\arcsec$ (235 au) above the envelope/disk mid-plane,
and the kinematics smoothly changes to that of rotating outflow at a larger $z$, 
we think that it is the expanding shell that produces such features.
The fact that the expanding motion of the outflow shell is only prominent at low $z$ 
can be explained by the curvature of the outflow shell.
At the base of the outflow cavity, along a line of sight toward the center of the outflow cavity,
the outflowing motion is mostly in the line-of-sight direction,
and also the column density of the molecule is higher due to the projection effect.
However, we cannot rule out the possibility that this is caused by infalling motion of the envelope.
At a height of $|z|=0.5\arcsec-1\arcsec$ ($100-200~\au$), it is possible that
the CCH emission traces the boundary between the outflow and the infalling envelope.
Currently, we lack enough spatial and velocity resolutions to disentangle
such spatial and kinematic changes.

\subsubsection{Angular Momentum in the Outflow}
\label{sec:angularmomentum}

To be more quantitative, we derive
some outflow properties in the following way.
In each panel of Figure \ref{fig:pvdiagram},
i.e. at each distance $z$ to the disk/envelope mid-plane,
we locate the emission peaks associated with the 
northern and southern outflow cavity walls,
with their line-of-sight velocities to be $\vlsrn(z)$ and $\vlsrs(z)$,
and their offsets to the designated outflow axis to be $R_\mathrm{N}(z)$ and $R_\mathrm{S}(z)$.
We define $R_\mathrm{out}(z)=[R_\mathrm{N}(z)-R_\mathrm{S}(z)]/2$ as 
the radius of the outflow cavity,
and $R_\mathrm{med}(z)=[R_\mathrm{N}(z)+R_\mathrm{S}(z)]/2$ as the distance
from the mid-point between the northern and southern cavity walls to the chosen outflow axis.
Assuming an edge-on inclination, the rotation velocity is
$\vrot(z)=[\vlsrs(z)-\vlsrn(z)]/2$, and the median line-of-sight velocity
$V_\mathrm{med}(z)=[\vlsrs(z)+\vlsrn(z)]/2$ should only reflect the outflowing velocity.
The specific angular momentum is then $j(z)=R_\mathrm{out}(z)\vrot(z)$.
Meanwhile, we calculate the velocity dispersions of the emission relative to $\vlsrn$ and $\vlsrs$ 
as the uncertainties in velocities,
and similarly we employ the widths of the outflow cavity wall at each $z$
as the uncertainties of the position offset determination.

\begin{figure}
\begin{center}
\includegraphics[width=0.5\textwidth]{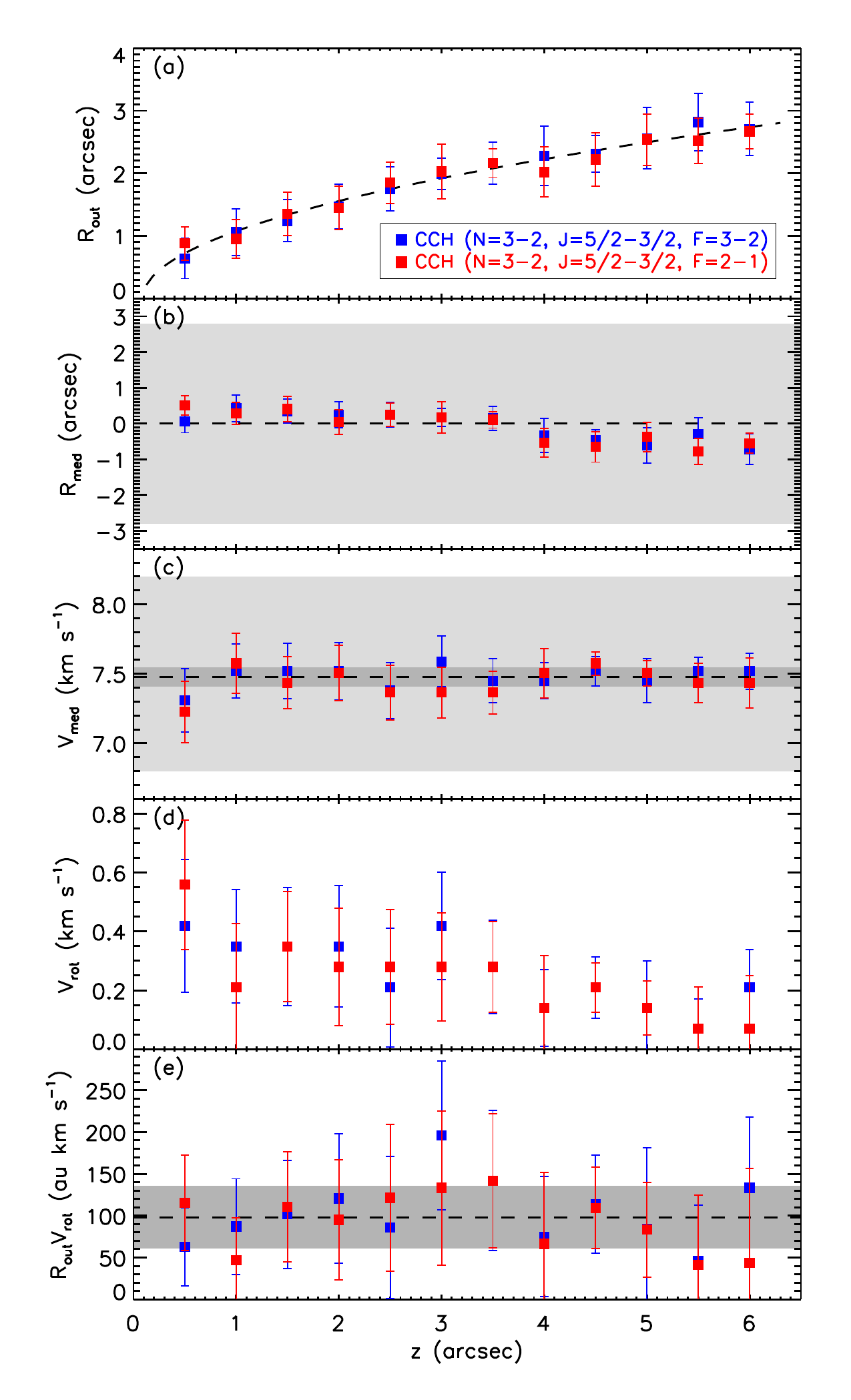}\\
\caption{The distribution of various properties of the outflow
derived from the CCH emission with the distance 
to the mid-plane of the envelope/disk structure.
{\bf (a):} The radius of the outflow cavity $R_\mathrm{out}$;
{\bf (b):} The deviation of the mid-point between the northern and southern cavity walls
from the chosen outflow axis $R_\mathrm{med}$.
{\bf (c):} The median line-of-sight velocity of the outflow $V_\mathrm{med}$; 
{\bf (d):} The rotation velocity of the outflow $\vrot$; 
{\bf (e):} The specific angular momentum of the outflow $R_\mathrm{out}\vrot$.
In all panels, blue and red symbols are from the $F=3-2$ and $F=2-1$
hyperfine lines of CCH, respectively.
In Panel (a), the dashed line is the parabolic fit to the outflow cavity shape.
In Panel (b), the shaded region indicates the maximum width of the outflow.
In Panel (c), the dashed line and dark shaded region indicate the mean value and standard deviation 
of the median line-of-sight velocities at $z>0.5\arcsec$.
The light shaded region indicates the velocity range in which the outflow is detected.
In Panel (e), the dashed line and shaded region indicate the mean value and standard deviation 
of the specific angular momentum.}
\label{fig:outflow}
\end{center}
\end{figure}

In Figure \ref{fig:outflow}, we show the profiles of $R_\mathrm{out}(z)$,
$R_\mathrm{med}(z)$, $\vrot(z)$, $V_\mathrm{med}(z)$, and $j(z)$.
The shape of the CCH outflow cavity can be described by a parabolic function
$z=0.79R^2+0.089$ with both $z$ and $R$ in arcsec.
In Panel (b), the distribution of $R_\mathrm{med}$ suggests that the centers of
the outflow cavity do not perfectly lie on the chosen outflow axis, although this deviation
is small compared with the width of the outflow cavity.
Such a deviation is not caused by the chosen position angle of the outflow axis,
but it rather indicates that
the axis of symmetry passes slightly to the north of the continuum peak,
which can be seen to some extent from the integrated maps shown in Figure \ref{fig:intmap}.
While this may be caused by a slight asymmetry in the distribution of dust emission 
in the disk, it is also possible that this is caused by a small outflow precession.
The median line-of-sight velocity of the outflow is quite flat except at $z=0.5\arcsec$,
with a mean value of about 7.5 $\kms$ and standard deviation
of 0.07 $\kms$ (excluding the data points at $z=0.5\arcsec$).
As discussed above, the nearly constant $V_\mathrm{med}$ indicates that 
either the outflow has a nearly constant forwarding velocity,
or the outflow is nearly on the plane of sky.
Panel (d) shows that the rotation velocity decreases with the distance to the central source
from about 0.5 $\kms$ at $z=0.5\arcsec$ (120 au) to about 0.1 $\kms$ at $z=6\arcsec$ (1400 au).
Combining the increasing $R_\mathrm{out}$ with $z$ and decreasing $\vrot$ with $z$,
the specific angular momentum $j=R_\mathrm{out}\vrot$ is largely flat.
The mean value of the specific angular momentum is about 100 $\aukms$
with a standard deviation of 40 $\aukms$.

\subsubsection{Launching Radius of the Outflow}
\label{sec:launching}

\begin{figure}
\begin{center}
\includegraphics[width=0.7\textwidth]{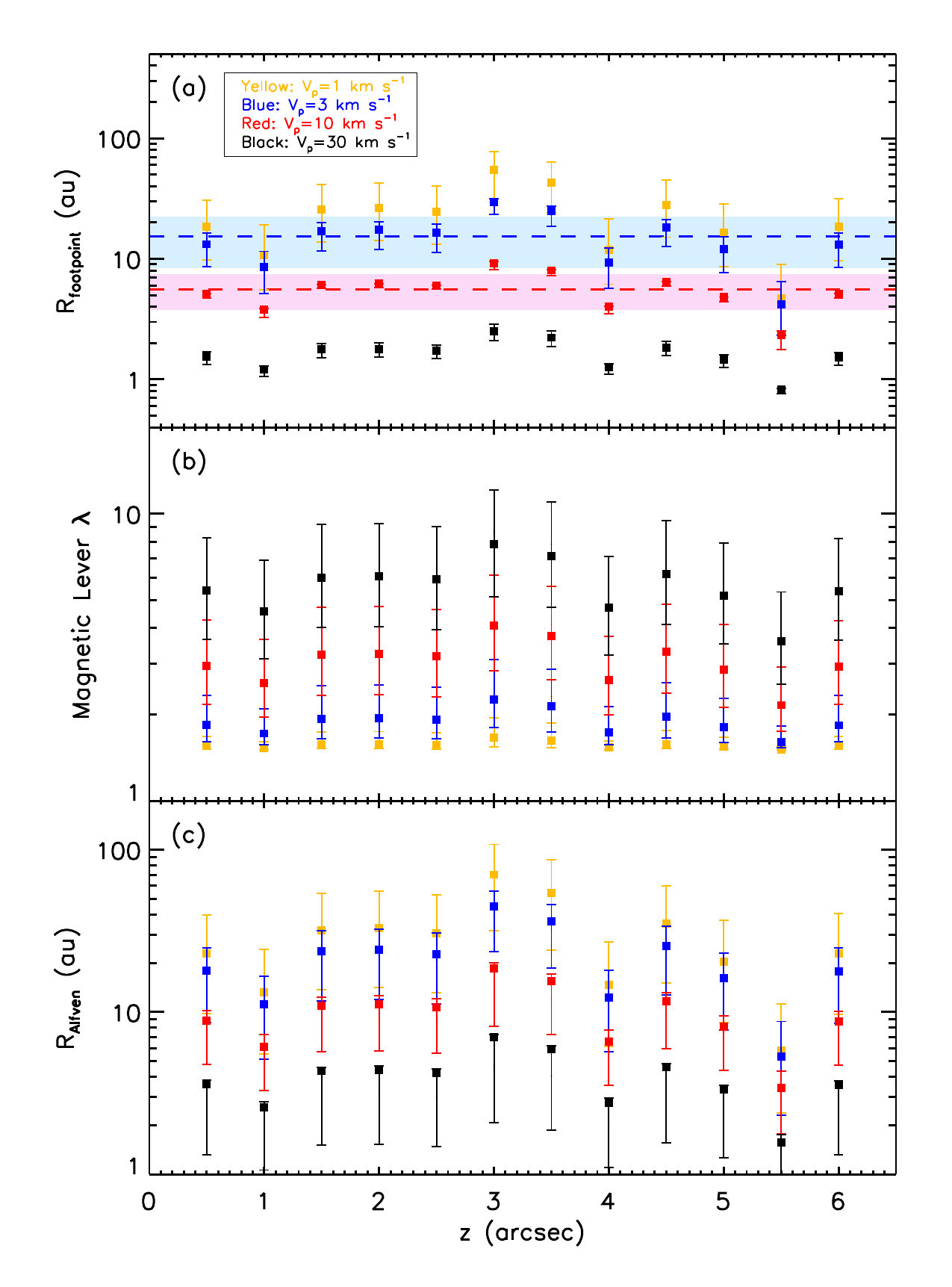}\\
\caption{The distributions of the derived launching radii $\rf$ (Panel a),
magnetic lever arm $\lambda$ (Panel b), and Alfv\'en radius $R_A$ (Panel c) of the outflow with the distance 
to the mid-plane of the envelope/disk structure.
Different poloidal velocities of the outflow are shown in different colors.
The data points show the fiducial values calculated with $m_*=0.2~M_\odot$,
while the error bars are calculated with $0.1~M_\odot < m_* < 0.4~M_\odot$
The dashed lines and shaded regions in Panel (a) show the mean values and standard deviations
of $\rf$ derived for the cases of $V_p=3$ (blue) and $10~\kms$ (red).}
\label{fig:outflow1}
\end{center}
\end{figure}

If the rotation signature that we detect originates from the rotation of a magneto-centrifugal wind
launched from the accretion disk,
the disk rotation rate at the outflow-launching region (and therefore the launching radius)
can be constrained by the outflow radius $R_\mathrm{out}$, the rotation velocity $\vrot$,
and the poloidal velocity of the outflow $V_p$, combined with the protostellar mass $m_*$,
following the method of \citet[]{Anderson03} (their Eq. 4).
Such a method is valid for general magneto-centrifugal winds launched from the disk, 
independent of the detailed configuration of magnetic field (disk winds or X-winds).
Panel (a) of Figure \ref{fig:outflow1} shows the derived launching radii of the outflow 
using the measured $R_\mathrm{out}$ and $\vrot$ (averaging the measurements 
from the two hyperfine emissions of CCH).

We adopt four values of $V_p$ (1, 3, 10 and 30 $\kms$) in the calculation. 
As discussed in \S\ref{sec:disk}, the eastern lobe may be blue-shifted by about 0.3 $\kms$
in relative to the central envelope structure, assuming a small
inclination angle of $i=5^\circ$, 
the forwarding velocity of the outflow is about 3.5 $\kms$.
The forwarding velocity becomes 0.7 $\kms$ if the inclination angle is $i=25^\circ$ (\citealt[]{SeguraCox16}).
An inclination higher than this is improbable for this source.
On the other hand, 
\citet[]{Koumpia16} found that, in $^{13}$CO $(J=2-1)$ emission, the peaks of the blue-shifted wing ($\vlsr=+5$ to $+6.5~\kms$)
and the red-shifted wing ($\vlsr=+8.5$ to $+11~\kms$) have a small offset. If we use half of the velocity differences between
blue and red wings as the projected outflow velocity, the forwarding velocity of the outflow is estimated to be about 15 $\kms$,
assuming $i=5^\circ$. However, the outflow rotation, which is not resolved in their observations, can also
contribute to the observed line width, and also the envelope material can contribute to the $^{13}$CO emission
at these low velocities. 
A much higher outflow velocity is unlikely (e.g. $\gtrsim30~\kms$ ), because even with a very small angle of inclination, 
it can make noticeable velocity differences between the eastern and western lobes.
Therefore, the chosen values of $V_p$ (1, 3, 10, and 30 $\kms$) cover a reasonable parameter space,
with $V_p=3-10~\kms$ being more probable, and $V_p=1$ and $30~\kms$ as lower and upper limits. 
The protostellar mass of this system is also uncertain. The luminosity of the system 
($L_\mathrm{bol}=0.7~L_\odot$) suggests that this is a protostar with a quite low mass.
In \S\ref{sec:disk}, we obtain a mass estimation of about $0.2~M_\odot$ 
from the dynamics of the envelope/disk system,
which we adopt as the fiducial value of the protostellar mass.
We also use of 0.1 and 0.4 $M_\odot$ as the lower and upper limits for $m_*$ in the calculation, 
which are 2 times lower or higher than the fiducial value.

Figure \ref{fig:outflow1}(a) shows the derived outflow launching radii using these $V_p$ and $m_*$.
The derived launching radii are largely flat with the distances to the central source $z$, 
i.e., the observed outflow at different distances is
launched from similar radii on the disk.
With the fiducial protostellar mass of $m_*=0.2~M_\odot$, the outflow launching radii
have a mean value of 15 au and a standard deviation of 7 au for the case of $V_p=3~\kms$,
and have a mean value of 5.6 au and a standard deviation of 1.8 au for the case of $V_p=10~\kms$.
A launching radius smaller than about 1 au is unlikely since it requires a high outflowing velocity.
On the other hand, a launching radius larger than about 50 au is also unlikely,
since it requires either a very low outflowing velocity or a much lower protostar mass.
Therefore, the most likely launching radii of the outflow are about 5 to 15 au,
with the uncertainties mostly due to the uncertainties in the outflow forwarding velocity.
The flat distribution of the derived outflow launching radii with $z$
is due to the flat angular momentum distribution 
and the fact that a constant $V_p$ is adopted for different $z$.
If the outflow is accelerating, e.g. from about $3~\kms$ at lower $z$ to about $10~\kms$ at higher $z$, 
the outflow at high $z$ should originate from a smaller radius (e.g. around 5 au)
in the disk. In contrast, while the outflow at low $z$ should originate from a larger radius (e.g. around 15 au),
which may indicate the growth of the outflow launching radius with time.
Such scenarios need future observations to distinguish.

The derived launching radii ($5 - 15$ au), even after considering the large uncertainties ($1 - 50$ au)
are consistent with disk wind models of magneto-centrifugal outflow (e.g. \citealt[]{Konigl00}).
In these models, the wind is launched from
a wide range of radii on the disk, as opposed to X-wind models (e.g. \citealt[]{Shu00}), 
in which the wind is launched from the very inner region of the disk (a few protostellar radii).
Such launching radii are also similar to the observations of other rotating wide and slow ($V_p<100~\kms$) protostellar outflows,
e.g. TMC1A (low mass Class I source, $\rf=5-25$ au; \citealt[]{Bjerkeli16}),
Orion Source I (massive protostar, $\rf=5-25$ au; \citealt[]{Hirota17}),
HH 212 (low mass Class 0 source, $\rf\lesssim40$ au according to \citealt[]{Tabone17},
and $\rf=2-3$ au according to \citealt[]{Lee18}),
and HH 46/47 (low mass Class I source, $\rf\sim60$ au; \citealt[]{Zhang16}).
In the case of HH 212, the rotation in the inner jet has also been reported by \citet[]{Lee17} with
a derived $\rf=0.05$ au for the jet. 
Small launching radii typically are found in searches of jet rotation,
ranging from a few $\times 10^{-2}$ to a few $\times 10^{0}$ au 
(e.g. \citealt[]{Lee09,Choi11,Chrysostomou08,Coffey07}).
Therefore, our results are consistent with a wide-angle disk wind launched from relatively large radii on the disk.

It is worth noting that the measured specific angular momentum $j=R_\mathrm{out}\vrot$ in the outflow 
is only the kinetic part of the total specific angular momentum. 
The total specific angular momentum, including additional magnetic contribution, 
should be conserved along a streamline and equals to $l=\Omega_0 R_A^2$,
where $\Omega_0$ is the angular velocity of the Keplerian disk at the launching radius $\rf$,
and $R_A$ is the Alfv\'en radius where the poloidal velocity of the outflow along the streamline
reaches the poloidal Alfv\'en speed.
The ratio between the total specific angular momentum of the outflow and
the specific angular momentum of the Keplerian disk at the launching radius is then
$\lambda=R_A^2/\rf^2$, which
is the magnetic lever arm that determines how much angular momentum is extracted from the disk.
Following Eq. 10 of \citet[]{Ferreira06}, we derive the magnetic lever arm $\lambda$
(Figure \ref{fig:outflow1}b)
and then Alfv\'en radius $R_A$ (Figure \ref{fig:outflow1}c), from the measured angular momentum, and
different assumed $V_p$ and $m_*$.
Previous studies show typical values of $\lambda$ of disk wind ranging from a few to 20 
(e.g. \citealt[]{Ferreira06,Tabone17}),
which favors the cases with $V_p=10~\kms$ or higher.
Such values of $\lambda$ give the Alfv\'en radii ranging from a few au to about 50 au.
Note that the derived Alfv\'en radii $R_A\ll R_\mathrm{out}$ (see Figure \ref{fig:outflow}, Panel a),
which suggests that at these radii most of the angular momentum has been transferred to the 
rotation from the magnetic field (\citealt[]{Ferreira06}). Therefore it is natural that the observed $j=R_\mathrm{out}\vrot$
shows a flat distribution with $z$.

\subsection{Connection with the Envelope/disk Structure}
\label{sec:disk}

\begin{figure}
\begin{center}
\includegraphics[width=0.8\textwidth]{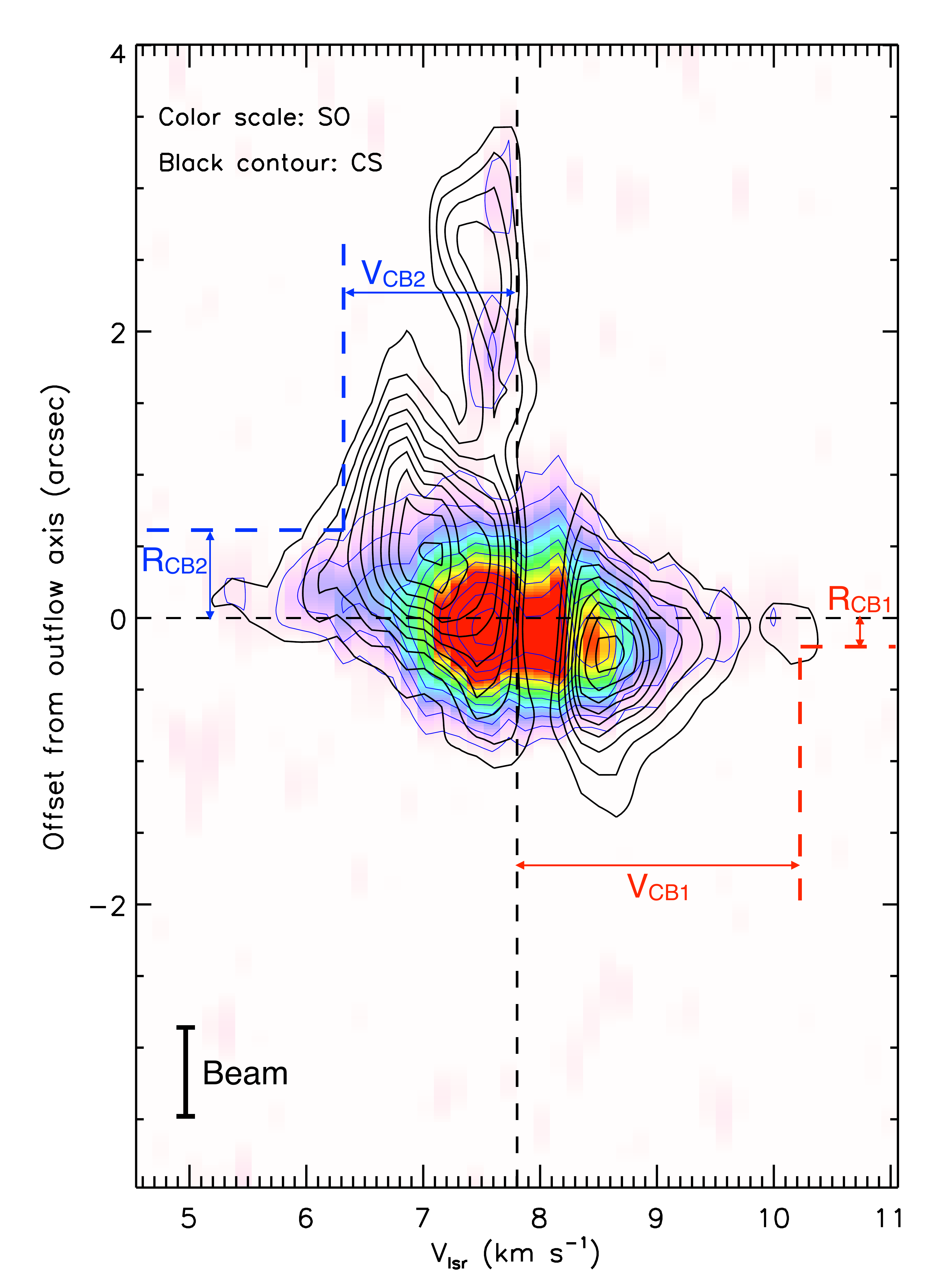}\\
\caption{The position-velocity diagrams of the $\CS$ emission (black contours)
and $\SO$ emission (color scale and blue contours) along a cut perpendicular to
the outflow axis and passing through the continuum peak, i.e. along the major axis of 
the continuum emission. The blue and red lines and labels are two estimations 
of the radius and velocity of the centrifugal barrier, based on the SO and CS emissions,
respectively (see the text for details).}
\label{fig:pvdiagram2}
\end{center}
\end{figure}

Figure \ref{fig:pvdiagram2} shows the PV diagrams of the $\CS$ and $\SO$ emissions
along the mid-plane of the envelope/disk.
While the SO emission is mostly concentrated within 1$\arcsec$ from the continuum peak,
the CS emission is broader toward the north and south.
Toward the north, there is extended CS emission further outside of $2\arcsec$
tracing the extended envelope.
In contrast, such an envelope component is not seen in the south.
These features may be due to the asymmetric distribution of envelope material or self-absorption.
It has been shown that, while CS mostly traces the envelope outward of the centrifugal barrier 
of the infalling envelope, 
SO is enhanced at the centrifugal barrier  (\citealt[]{Sakai14,Oya15})
by accretion shocks liberating SO from dust grain surfaces to the gas phase. 
Therefore, it is natural that the SO emission is more confined around the protostar than CS.
In such a case, the highest velocity seen in CS and its position could measure
the velocity and radius of the centrifugal barrier.
However, in some cases, CS can survive somewhat inside the centrifugal barrier (e.g. \citealt[]{Oya17})
and also probe higher velocities in the disk.
As Figure \ref{fig:pvdiagram2} shows, 
there is not enough spatial resolution
to clearly distinguish these two scenarios.
Even if the SO traces a higher velocity of the disk Keplerian rotation,
it can be smeared out and mixed with the rotation of the inner envelope.
On the other hand, since SO emission/abundance is enhanced 
due to the accretion shock around the centrifugal barrier,
the radius of the SO enhanced emission can also suggest the radius of the centrifugal barrier.

Therefore, we estimate two sets of centrifugal barrier properties.
In the first method, assuming that CS only traces the envelope outside of the centrifugal barrier,
we use the highest velocity seen in CS and its position 
as the velocity and radius of the centrifugal barrier,
which gives $V_\mathrm{CB1}=2.5~\kms$ and $R_\mathrm{CB1}=0.2\arcsec$, i.e. 50 au
(the red lines and labels in Figure \ref{fig:pvdiagram2}).
In the second method, we use the outer radius of the SO emission and the maximum velocity at that radius as
the velocity and radius of the centrifugal barrier (see \citealt[]{Sakai14a}),
assuming SO is only enhanced around the centrifugal barrier and inside.
In this case, we obtain $V_\mathrm{CB2}=1.5~\kms$ and $R_\mathrm{CB2}=0.6\arcsec$, i.e. 140 au
(the blue lines and labels in Figure \ref{fig:pvdiagram2}).
Following \citet[]{Sakai14}, we then estimate the protostellar mass 
to be $M_*=V_\mathrm{CB}^2 R_\mathrm{CB}/(2G)\approx0.18~M_\odot$ from
both sets of estimations.

As discussed in \S\ref{sec:launching}, the launching radii of the outflow are estimated 
to be $5 - 15~\au$, which is smaller than the radius of the centrifugal barrier 
obtained above ($50 - 140~\au$).
This is reasonable since the centrifugal barrier can be
considered as the outer radius of the accretion disk.
Some studies have indicated that the centrifugal barrier may play a role in 
the outflow launching (\citealt[]{Sakai17,Alves17}).
However, it is not clear whether the centrifugal 
barrier is directly related to the outflow launching in this source,
considering their different radii.
The specific angular momentum at the centrifugal barrier is estimated to be $125 - 210~\aukms$, 
which is comparable to the mean specific angular momentum of the 
outflow (about $100~\aukms$). 
However, since the location and velocity of the centrifugal barrier 
are only roughly estimated without resolving spatially, 
they would have large uncertainties. 
The possibility of a smaller centrifugal barrier radius which is closer to the outflow launching radii
cannot be ruled out. 
\citet[]{SeguraCox16} estimated a disk radius of about 30 au for this source
by fitting 8 mm continuum emission,
which is closer to the outflow launching radii estimated here.
Apparently, high angular resolution observations are needed.

Finally, we note that the median velocities of the envelope traced by CS and SO emissions 
are about $\vlsr=7.8~\kms$, which is $0.3~\kms$ offset from
the median velocities of $\vlsr=7.5~\kms$ of the eastern lobe (see Panel c of Figure \ref{fig:outflow}),
indicating that the eastern lobe is indeed slightly blue-shifted. 

\section{Conclusions}
\label{sec:conclusion}

We present ALMA observations of the NGC 1333 IRAS 4C outflow,
including CCH, CS and SO emission.
Our main conclusions are as follows.

1) The outflow cavity structure is detected and resolved in the CCH and CS line emission.
This is the first time that the molecular outflow associated with this source is clearly detected.
The morphology of this molecular outflow coincides with the cavity structure
seen in MIR emissions.
While the CCH emission appears to mostly trace the outflow cavity structure,
the CS emission traces the envelope.
The morphology and kinematics of the molecular emission
are consistent with an almost edge-on inclination of the disk, as previously speculated.

2) The observed velocity pattern of the CCH emission is 
dominated by the rotation of the outflow cavity walls around its axis.
The contribution of the outward motion of the outflow
is very minor due to the nearly edge-on inclination.
The rotation is detected from about 0.5$\arcsec$ (120 au) above the envelope/disk mid-plane
up to the full extension of the CCH outflow at 6$\arcsec$ (1400 au) from the central source.
The velocity pattern is highly symmetric with respect to the outflow axis.
The measured rotation velocity is about 0.5 $\kms$ at 0.5$\arcsec$ above the disk,
and smoothly decreases to about 0.1 $\kms$ at 6$\arcsec$ above the disk.
Because of the parabolic shape of the outflow cavity wall, the angular momentum
of the outflow is nearly flat with the distance to the protostar.
The mean specific angular momentum in the outflow is estimated to be about 100 $\aukms$.

3) From the derived angular momentum in the outflow, by assuming reasonable forwarding velocity
of the outflow and protostellar mass, 
we infer the most likely launching radii of the outflow
to be $5-15$ au. The derived launching radii are similar to those
derived from observations of rotation in other wide slow molecular outflows.
Such a range of launching radii is consistent with the picture of a disk wind in which the outflow
is launched over a range of radii of the disk, as opposed to an X-wind in which the outflow is launched
from the innermost region of the disk (typically a few protostellar radii).

4) From the CS and SO emission which traces the envelope and maybe disk, we roughly estimate a radius of the centrifugal barrier
of about $50-140$ au and a specific angular momentum of about $120-210~\aukms$ at the centrifugal barrier.
Our current observations do not have enough resolution to resolve the transition from the envelope
to the disk and clearly locate or even resolve the centrifugal barrier structure.
Therefore the above estimated radius and specific angular momentum are 
highly uncertain. Thus, the higher angular resolution observations are needed in 
order to study connections between the outflow launching and the centrifugal 
barrier.

\acknowledgements
This paper makes use of the following ALMA data: ADS/JAO.ALMA\#2016.1.01501.S. 
ALMA is a partnership of ESO (representing its member states), 
NSF (USA) and NINS (Japan), together with NRC (Canada), 
MOST and ASIAA (Taiwan), and KASI (Republic of Korea), 
in cooperation with the Republic of Chile. 
The Joint ALMA Observatory is operated by ESO, AUI/NRAO and NAOJ.
This project is supported by Grant-in-Aids from Ministry of Education, Culture, Sports, Science, 
and Technologies of Japan (25108005 and 16H03964). 
The authors acknowledge financial support by JSPS and MAEE under the Japan/France integrated action 
programme. 
Y.Z. acknowledges support from RIKEN Special Postdoctoral Researcher Program.

\software{CASA (\citealt[]{McMullin07})}

\clearpage

\begin{table} 
\scriptsize
\begin{center}
\caption{Parameters of the Observed Lines\footnote{Information taken from the CDMS database (\citealt[]{Muller05})} \label{tab:lines}}
\begin{tabular}{lccccc}
\hline
\hline
Molecule & Transition & Frequency (GHz) & $E_u/k$ (K) & $S\mu^2$ (D$^2$) & Synthesized Beam \\
\hline
CCH & $N=3-2,~J=5/2-3/2,~F=2-1$ & 262.0674690 & 25.2 & 1.067 & $0.64\arcsec\times0.41\arcsec$  ($\pa=-6.4^\circ$)\\
CCH & $N=3-2,~J=5/2-3/2,~F=3-2$ & 262.0649860 & 25.2 & 1.633 & $0.64\arcsec\times0.41\arcsec$  ($\pa=-6.4^\circ$)\\
CS & $J=5-4$ & 244.9355565 & 35.3 & 19.17 & $0.68\arcsec\times0.43\arcsec$  ($\pa=-6.8^\circ$)\\
SO & $J_N=6_6-5_5$ & 258.2558259 & 56.5 & 13.74 & $0.64\arcsec\times0.41\arcsec$  ($\pa=-7.0^\circ$)\\
\hline
\end{tabular}
\end{center}
\end{table}

\clearpage

\clearpage

\clearpage

\clearpage

\clearpage

\clearpage

\clearpage

\clearpage

\end{document}